\documentclass[12pt, a4paper]{article}

\usepackage{graphicx}
\usepackage{bmpsize}
\usepackage{amsmath}
\usepackage{amsfonts}
\usepackage{amssymb}
\usepackage{latexsym}
\usepackage{graphicx}
\usepackage{color}
\usepackage{mathrsfs}
\usepackage{slashed}
\usepackage{soul}
\usepackage{array}
\usepackage{placeins}
\usepackage{hyperref}

\usepackage{enumerate} 
\usepackage{multirow}
\usepackage[table]{xcolor}
\usepackage{colortbl}
\usepackage{tikz-feynman}
\tikzfeynmanset{compat=1.0.0}

\usepackage{subcaption}

\usepackage{float}
\usepackage{braket}
\usepackage{xfrac}
\usepackage
[
version=4,
math-greek=default,
text-greek=default,
]{mhchem
}
\usepackage{booktabs}  

\usepackage[text={7in,10in}]{geometry}

\usepackage{comment}
\includecomment{toexclude} % you can name the comment as you wish
\excludecomment{addsections} % you can name the comment as you wish

\usepackage[
  output-decimal-marker=.,
  separate-uncertainty=true,
  per-mode=symbol-or-fraction,
]{siunitx}
\DeclareSIUnit\parsec{pc}
\DeclareSIUnit\erg{erg}

\usepackage{color}

\usepackage{slashed}
\usepackage{xfrac}

\numberwithin{equation}{section}

\newcommand{\lag}{\mathcal L}

\newcommand{\hc}{\mathrm{h.c.}}
\newcommand{\tr}{\mathrm{Tr}}
\newcommand{\m}[1]{\mu_{#1}}
\newcommand{\bn}{\mathcal{B}}
\newcommand{\lp}{\mathcal{L}}
\newcommand{\uell}{U(1)_{\mathcal{L}}}
\newcommand{\ub}{U(1)_{\mathcal{B}}}
\newcommand{\yd}[1]{Y_{\Delta_{#1}}}

\begin{document}

\begin{center}
\hfill\textit{DO-TH 22/02}\\
\vspace{1cm}
{\Huge
Leptogenesis in Majoron Models without Domain Walls
}
\\ [2.5cm]
{\large{\textsc{ 
Tim Brune\footnote{\textsl{tim.brune@tu-dortmund.de}}
}}}
\\[1cm]

\large{\textit{
Fakult\"at f\"ur Physik, Technische Universit\"at Dortmund,\\
44221 Dortmund, Germany
}}
\\ [2 cm]
{ \large{\textrm{
Abstract 
}}}
\\ [1.5cm]
\end{center}
The emergence of domain walls is a well-known problem in Majoron models for neutrino mass generation. Here, we present extensions of the Majoron model by right-handed doublets and triplets that prevent domain walls from arising. These extensions are highly interesting in the context of leptogenesis as they impact the conversion of a lepton asymmetry to a baryon asymmetry by sphaleron processes.

\def\thefootnote{\arabic{footnote}}
\setcounter{footnote}{0}
\pagestyle{empty}

\newpage
\pagestyle{plain}
\setcounter{page}{1}

\section{Introduction}
By now, the existence of at least two nonvanishing neutrino masses is well established by means of the observation of neutrino oscillations \cite{NO_kamland}-\cite{NO_superkamiokande}. Nevertheless, the Standard Model (SM) of particle physics does not provide a mechanism for their generation, let alone an explanation for their smallness compared to the masses of the other SM particles. 
On the cosmological side, the evidence for the existence of non-baryonic dark matter (DM) in the Universe, the flatness, homogeneity and isotropy of the Universe just as the observed baryon asymmetry of the Universe indicate the need for physics models beyond the SM and hint at the existence of new particles. \\
The singlet Majoron model \cite{majoron_chikashige}-\cite{majoron_gelmini} is a very compelling extension of the SM, as it requires minimal physics beyond
the SM and, nevertheless, can address all of the previously stated problems. 
It is based on the spontaneous symmetry breaking (SSB) of a global $U(1)_{\bn-\lp}$, an accidental symmetry of the SM, at the seesaw scale. As a consequence of SSB, a Goldstone boson, called the Majoron, arises, and small neutrino masses are generated by the seesaw mechanism. 
If the Majoron obtains a mass of the order of MeV due to the existence of explicit breaking terms and consequently becomes a pseudo-Goldstone boson, a Majoron relic density in accordance with the DM relic density can be produced via freeze-in \cite{brune_paes}-\cite{frigerio}. Moreover, lepton-number-violating decays of right-handed neutrinos can create a lepton asymmetry which is subsequently transferred to a baryon asymmetry by sphaleron processes \cite{leptogenesis}-\cite{2014}, thereby accounting for the baryon asymmetry.\\
However, as pointed out in \cite{dw1}, domain walls \cite{vilenkin} form in the singlet Majoron model due to the interplay of nonperturbative instanton effects and SSB at the seesaw scale. 
More precisely, the instanton processes break the inital $\uell$ symmetry to a residual $Z_3$, while SSB at the seesaw scale breaks the $\uell$ symmetry to a residual $Z_2$, implying a mismatch of discrete symmetries and resulting in the formation of domain walls. The appearance of domain walls is highly undesired, as such topological defects typically dominate the energy density of the Universe, contrary to observations. 
In this paper, we discuss several models that avoid the appearance of domain walls by introducing new particles, right-handed doublets and triplets. \\
Besides addressing the problematic formation of domain walls, the extension of the Majoron model has interesting consequences for the leptogenesis scenario taking place in such models. The same mechanism that prevents the emergence of domain walls changes the amount of lepton number that is violated by sphaleron processes, thus affecting the amount of baryon asymmetry that can be generated from an initial lepton asymmetry. Assuming that a lepton asymmetry has been generated, for example by lepton-number-violating decays of right-handed neutrinos, we calculate the amount that is transferred to a baryon asymmetry by sphaleron processes. \\
This paper is organized as follows: In Sec. \ref{sec:majoron}, we revisit the Majoron model and discuss the leptogenesis mechanism in these models. After that, in Sec. \ref{sec:anomalies}, we give a brief introduction to anomalies and instanton processes. Finally in Sec. \ref{sec:nodw}, we give examples for models that are safe from the occurence of domain walls and discuss their impact on the leptogenesis mechanism. 

\section{The Majoron Model}
\label{sec:majoron}
\subsection{Model Setup}
\label{sec:majoron_setup}
In the singlet Majoron model, the SM is extended by a singlet complex scalar $\sigma$ with a vacuum expectation value (VEV) $f$ and three right-handed neutrinos $N_R$, transforming as 
\begin{align}
  \sigma \sim (1,1,0)_{-2}\,,\qquad\qquad  N_R \sim (1,1,0)_1 \,,
\end{align}
under $(SU(3)_C\times SU(2)_L\times U(1)_Y)_{\mathcal{L}}$ where the index $\mathcal{L}$ denotes the lepton number. The $U(1)_{\mathcal L}$ invariant Lagrangian coupling $N_R$ to $\sigma$ and the Higgs doublet $H$, 
\begin{align}
   H = \begin{pmatrix} \phi^+ \\ \phi^0 \end{pmatrix} \,,
\end{align} 
is given by
\begin{align}
  \lag_{yuk}^{new} &= -y_{H_{ij}}^\nu \overline{L_{L_i}}\tilde HN_{R_j}- \frac{1}{2} g_{N_{ij}} \overline{N_{R_i}^c}N_{R_j}\sigma  + \hc\,.
\end{align}
Moreover, the existence of the scalar $\sigma$ gives rise to new terms in the $U(1)_{\mathcal L}$ invariant scalar potential, 
\begin{align}
  V(H, \sigma) = -\mu_\sigma^2 |\sigma|^2 + \lambda_\sigma^2 |\sigma|^4 -\mu_H^2 |H|^2 + \lambda_H^2 |H|^4 +2\lambda_{mix}|\sigma|^2|H|^2 \,.
\end{align}
The $\uell$ symmetry is spontaneously broken at the seesaw scale $f\sim \SI{e9}{\giga\electronvolt}$ and consequently, $\sigma$ can be expanded around its ground state as 
\begin{align}
  \sigma &= \frac{1}{\sqrt2}(f + \sigma^0 + \mathrm{i}J) \,,
\end{align}
where $J$ is the CP-odd Majoron and $\sigma^0$ is its CP-even partner. 
After the electroweak phase transition (EWPT), $H$ can be written as 
\begin{align}
  H &=\frac{1}{\sqrt 2}\begin{pmatrix} 0 \\ v+ h^0 \end{pmatrix}\,,
\end{align}
where $h^0$ is the Higgs boson and $v$ is the VEV. 
The subsequent symmetry breakings give rise to a nondiaginal mass matrix for the neutrinos with a Majorana mass $M_R = \frac{fg_N}{\sqrt2}$ and a Dirac mass $m_D = \frac{vy_H^\nu}{\sqrt2}$. Diagonalization in the seesaw limit $M_R \gg m_D$ yields the mass eigenstates of the heavy Majorana neutrinos, defined as $N := N_R + N_R^c$, as $m_h \propto M_R$ and of the light neutrinos as 
$m_l \propto -\frac{m_Dm_D^T}{M_R}$.\\
Moreover, we assume that a Majoron mass of the order of $\si{\mega\electronvolt}$ is generated by a radiatively induced term \cite{frigerio, neutrinolines}
\begin{align}
  \lag_{H,\sigma} =  \lambda_{hJ}J^2 |H|^2 + \hc\,,
\end{align}
thereby establishing the Majoron as a DM candidate \cite{brune_paes}-\cite{frigerio}. \\
{On tree-level, the Majoron couples only to the mass eigenstates of the neutrinos with a coupling $\mathcal{O}(\frac{m_i}{f})$, where $m_i \lesssim \SI{e-2}{\electronvolt}$. For a seesaw scale $f \sim \SI{e9}{\giga\electronvolt}$, this tiny coupling ensures that the Majoron can easily be longlived enough to make up the DM of the Universe. As the decays of the Majoron are to neutrino mass eigenstates, the neutrinos do not oscillate, resulting in a monochromatic neutrino flux \cite{neutrinolines}.  }\\
% In the given model, the baryon asymmetry generated from an initial lepton asymmetry is given by 
% \begin{align}
%   Y_{\Delta \mathcal B} = Y_{\Delta(\mathcal B- \mathcal L)} \begin{cases} \frac{28}{79} \qquad T>T_{EWPT}\\ \frac{12}{37}\qquad T<T_{EWPT}\,. \end{cases}
% \end{align}
In the scenario explained above, the global $U(1)_{\mathcal L}$ symmetry is broken down to a residual $Z_2$ symmetry which can be easily seen when $\sigma$ is written in the radial representation, ${\sigma\propto(f+\sigma^0)\mathrm{e}^{\mathrm{i}\frac{J}{f}}}$. Before SSB, the Lagrangian is invariant under $U(1)_{\mathcal L}$ transformations given by 
\begin{align}
(L, N_R) &\to (L,N_R)\mathrm{e}^{-\mathrm{i}\alpha_{\lp}}\,, \\
(\overline{L}, \overline{N_R}) &\to (\overline{L},\overline{N_R})\mathrm{e}^{\mathrm{i}\alpha_{\lp}}\,, \\
\sigma &\to \sigma\mathrm{e}^{2\mathrm{i}\alpha_{\lp}}\,. 
\end{align}
After SSB, a $Z_2$ subgroup of $U(1)_\lp$ remains unbroken where
\begin{align}
(L, N_R) &\to (L,N_R)\mathrm{e}^{-\mathrm{i}k \pi } = \pm (L,N_R)\,, \\
(\overline{L},\overline{N_R}) &\to (\overline{L},\overline{N_R})\mathrm{e}^{\mathrm{i}k \pi }= \pm (\overline{L},\overline{N_R})\,, \\
J &\to J + 2\pi k f\,,
\end{align}
with $k= 0,1$. 
Consequently, the vacuum can be written as 
\begin{align}
|\braket{\sigma}| \propto f\mathrm{e}^{\mathrm{i}\delta}\mathrm{e}^{2\mathrm{i}k\pi}\,
\end{align}
where $\delta$ is an arbitrary phase. 
%We note that in the case given, we have one vacuum with an arbitrary phase (domain walls and strings?). \\
We want to stress that we do not take effects of gravitational instantons into account, thus unlike in i.e. axion models, the Majoron does not acquire an effective potential which may also result in the appearance of domain walls \cite{dw2}-\cite{Wilczek:1977pj}. 

\subsection{Leptogenesis}
In the conventional leptogenesis scenario, a $\lp$ asymmetry (and thereby a $\mathcal B-\mathcal L$ asymmetry where $\bn$ denotes baryon number) is generated by $\lp$-violating out-of-equilibrium decays of heavy Majorana neutrinos, $N \to \ell H,\overline\ell H^*$. 
This asymmetry is subsequently transferred to a $\bn$ asymmetry by sphaleron processes (see Sec. \ref{sec:anomalies}). However, it is possible that the asymmetry is washed out due to inverse decays, $\ell H,\overline\ell H^* \to N$, and $\lp =2$ violating scattering processes $H \ell \leftrightarrow H^*\overline{\ell}$. \\
Because of the presence of heavy Majorana neutrinos, the Majoron model offers the basic framework for leptogenesis models. 
The relevant terms in the Lagrangian are given by 
\begin{align}
\begin{split}
  \lag = &-(y_{H_{ij}}^\nu \overline{L_{L_i}}\tilde HN +\hc) - \frac{1}{2}M_N \overline NN -\frac{\mathrm{i}}{2\sqrt{2}} g_N \overline N \gamma_5 N J -\frac{\mathrm{1}}{2\sqrt{2}} g_Nh\sigma^0 \overline NN \\&- \lambda_\sigma^2f\sigma^0J^2- \lambda_{mix}f\sigma^0|H|^2- \lambda_{mix}f{\sigma^0}^3 \,,
\end{split}
\end{align}
where the last five terms arise exclusively in the Majoron model. They result in additional scattering processes, $NN\leftrightarrow JJ,HH^\star, \sigma^0\sigma^0$. While these processes could result in a larger washout compared to the conventional leptogenesis mechanism, they could also significantly enhance the final $\bn-\lp$ asymmetry \cite{2014,2011}.

\section{Anomalies and Instantons in the SM}
\label{sec:anomalies}
In this section, we review the $\mathcal{B}$ and $\mathcal{L}$ anomalies of the standard model and their conjunction with instanton and sphaleron processes. A more comprehensive introduction to anomalies can be found in \cite{bilallectures} while instantons are extensively reviewed in \cite{usesofinstantons}.
\subsection{Anomalies}
In the SM, baryon number $\ub$ and lepton number $\uell$ are accidental global symmetries that are conserved to all orders of perturbation theory. The corresponding charges are defined as
\begin{align}
  Q_{\mathcal{B,L}} = \int \mathrm{d}^3 x J_{\mathcal{B,L}}^0\,,
\end{align}
where the baryon and lepton current are given by
\begin{align}
  J_{\bn}^\mu &= \frac{1}{3}\left[ \overline q\gamma^\mu q + \overline{u_R}\gamma^\mu u_R+ \overline{d_R}\gamma^\mu d_R  \right]\,,\\
  J_{\lp}^\mu &= \overline \ell\gamma^\mu \ell + \overline{e_R}\gamma^\mu e_R\,,
\end{align}
respectively. 
Moreover, $\ell$ and $q$ are the lepton and quark doublets, $e_R, u_R, d_R$ are the lepton and quark singlets and generation and color indices are implicit.
However, both symmetries are anomalous symmetries with respect to $SU(2)_L$ with anomalies given by \cite{adler,belljackiw}
\begin{align}
  \partial_\mu J^{\mu}_{\mathcal{B,L}} = \frac{g^2A_{\mathcal{B,L}}}{32\pi^2}F_{\mu\nu}\tilde F^{\mu\nu}\,. \label{anomaly}
\end{align}
Here, $F_{\mu\nu}$ is the field strength tensor of the $W$-boson, 
\begin{align}
  F_{\mu\nu}^a = \partial_\mu W_\nu^a - \partial_\nu W_\mu^a + g f^{abc} W_\mu^bW_\nu^c\,,\quad a,b,c= 1,2,3\,,
\end{align}
and $g$ is the $SU(2)_L$ gauge coupling. The anomaly factor $\mathcal{A}_q$ for a general $\left[SU(2)_L\right]^2\times U(1)_q$ anomaly is given by 
\begin{align}
  \mathcal{A}_q = \sum_{\mathcal{R}} \left[ \sum_{L}2qT(\mathcal R)- \sum_{R}2qT(\mathcal R) \right]\,. \label{eq:aq}
  % \mathcal{A}_q = \sum_{\mathcal{R}} \left[ \sum_{L}q\tr_{\mathcal{R}} \{ T^b, T^c \}- \sum_{R}q\tr_{\mathcal{R}} \{ T^b, T^c \} \right]\,, \label{eq:aq}
\end{align}
Above, $q$ is the $U(1)_q$-charge and $T(\mathcal R)$ is the index of the representation under which a left-handed ($L$) or right-handed ($R$) particle transforms. For singlets, doublets and triplets, we have $T(1) = 0, T(2) = \frac{1}{2}$ and $T(3) =2$, respectively. \\
In the SM, the $[SU(2)_L]^2\times U(1)_{\mathcal{L,B}}$ anomaly factors are given by 
\begin{align}
  \mathcal{A}_{\mathcal{L}} &=\sum_{\mathcal{R}}  N_L(\mathcal{R})\times \lp_L(\mathcal{R})\times 2T_L(\mathcal{R}) = 3\times1\times 2\frac{1}{2} = 3\,, \\
  \mathcal{A}_{\mathcal{B}} &=\sum_{\mathcal{R}}  N_L(\mathcal{R})\times C_L(\mathcal{R})\times \bn_L(\mathcal{R})\times 2T_L(\mathcal{R}) = 3\times3\times\frac{1}{3}\times 2\frac{1}{2} = 3 \,,
\end{align}
where $N_L$ is the number of generations of left-handed particles, $C_L$ is the color multiplicity and $\lp_L$ and $\bn_L$ are the lepton and baryon numbers of the respective left-handed particle. As the SM contains only left-handed particles with nontrivial $SU(2)_L$ quantum numbers, right-handed particles do not contribute to the anomalies.
\subsection{Instantons and Sphalerons}
The electroweak theory possesses a nontrivial vacuum structure where the gauge-invariant ground state, the so-called theta vacuum, is given by a superposition of topological winding sectors $n$ as
\begin{align}
\ket{\theta} = \mathcal N \sum_n \mathrm{e}^{\mathrm{i}n\theta}, \quad\theta = [0,2\pi)\,.
\end{align}
The effect of the theta vacuum can be incorporated in the theory as an additional term in the action, 
\begin{align}
  S_\theta = \theta Q\,,
\end{align}
where $Q$ is the topological charge, 
\begin{align}
  \mathcal{Q} = \int \mathrm{d}^4x \partial_\mu K^\mu = n(\infty) - n(-\infty) = \frac{g^2}{32\pi^2}\int F_{\mu\nu}\tilde F^{\mu\nu}\,.
\end{align}
From the equation above, the relation between the topological charge $Q$ and the anomaly given in \eqref{anomaly} is apparent. \\
% We can relate the anomaly given in \eqref{anomaly} to the so called topological charge $\mathcal{Q}$, given by 
% \begin{align}
%   \mathcal{Q} = \int \mathrm{d}^4x \partial_\mu K^\mu = K(\infty) - K(-\infty) = \frac{g^2}{32\pi^2}\int F_{\mu\nu}\tilde F^{\mu\nu}\,,
% \end{align}
% where $K$ is the Chern-Simons charge or winding number. 
At zero temperature, instanton processes interpolate between neighboring winding sectors, i.e. $\mathcal{Q} = 1$. Comparing with \eqref{anomaly}, we observe that the transitions between the neighboring sectors violate $\bn$ and $\lp$ by $3$ units while conserving $\bn-\lp$. Consequently, the $U(1)_{\lp,\bn}$ symmetries are broken \cite{thooft}. However, a residual $Z_3$ symmetry of each continous symmetry remains unbroken. 
This can be seen when the generating functional, 
\begin{align}
   Z = \int\mathcal{D}\psi \mathcal{D} \overline\psi \mathrm{e}^{\mathrm{i} S}\mathrm{e}^{\mathrm{i} S_\theta} \,,
\end{align}
is considered where $S = \int \mathrm d^4 x \mathrm i \overline \psi \gamma^\mu \partial_\mu \psi$ is the action for a Dirac fermion and we explicitly included $S_\theta$. 
Under a transformation
\begin{align}
  \psi \to \mathrm{e}^{\mathrm i \alpha}\psi\,, \label{eq:rot}
\end{align}
the action is invariant. However, the measure $\mathcal D\psi \mathcal D \overline\psi$ is not invariant and a term is induced which can be written as a contribution to the action as
\begin{align}
  Z \to \int\mathcal D\psi \mathcal D \overline\psi \mathrm{e}^{\mathrm{i} S}\mathrm{e}^{\mathrm{i} S_\theta}\mathrm{e}^{-\mathrm{i}\alpha \int \mathrm d^4 x\partial_\mu J^\mu_q} = \int\mathcal D\psi \mathcal D \overline\psi \mathrm{e}^{\mathrm{i} S}\mathrm{e}^{\mathrm{i} S_\theta}\mathrm{e}^{-\mathrm{i}\alpha \mathcal A_q Q} = \int\mathcal D\psi \mathcal D \overline\psi \mathrm{e}^{\mathrm{i} S}\mathrm{e}^{\mathrm{i} (\theta - \mathrm{i}\alpha \mathcal A_q) Q} 
  % Z \to \int\mathcal D\psi \mathcal D \overline\psi \exp{\left[\mathrm{i}S  -\mathrm{i}\alpha \int \mathrm d^4 x\partial_\mu J^\mu_q\right]} = \int\mathcal D\psi \mathcal D \overline\psi \exp{\left[\mathrm{i}S  -\mathrm{i}\alpha \mathcal A_q\right]}\,.
\end{align}
If \eqref{eq:rot} corresponds to a $U(1)_{\lp}$ transformation, we have
\begin{align}
 \int\mathcal D\psi \mathcal D \overline\psi \mathrm{e}^{\mathrm{i} S}\mathrm{e}^{\mathrm{i} (\theta - \mathrm{i}3\alpha \mathcal ) }\,
\end{align}
and we observe that the $U(1)_{\lp}$ symmtery is explicitly broken to a residual $Z_3$ symmetry generated by $\alpha=\frac{2}{3}k\pi\,, k = 0,1,2$.\\
The instanton processes can be interpreted as a transition between the winding sectors via tunneling and as a consequence, they are highly suppressed by a factor $\mathrm{e}^{-\frac{4\pi}{\alpha_W}} \sim \num{e-150}$, where $\alpha_W = \frac{g^2}{4\pi}$. At high energies, $\SI{e12}{\giga\electronvolt} > T > \SI{e2}{\giga\electronvolt}$, so called sphaleron processes can overcome the barrier, resulting in a sizable $\bn+\lp$ violation and the explicit breaking of the $U(1)_{\lp,\bn}$ to $Z_3$ symmetries. 
If an initial $\bn-\lp$ asymmetry exists, for example due to the lepton-number-violating decays of right-handed neutrinos at high energies, fast sphaleron processes can transfer the $\bn-\lp$ asymmetry to a $\bn$ asymmetry. \\
The amount of asymmetry that is transferred can be calculated by considering the relation between the chemical potentials of SM particles induced by sphaleron processes being in thermal equilibrium, 
\begin{align}
  2\m{d_L} + \m{u_L} + \m{\nu_L} = 0\,.
\end{align}
Fast electroweak and Yukawa interactions result in additional relations between the chemical potentials, 
\begin{align}
  \m{W^-} =& \m{\phi^0} + \m{\phi^-} = \m{d_L} - \m{u_L} = \m{e_L}-\m{\nu_L} \,,\label{eq:chemequ1}\\
  \m{\phi^0} =& \m{u_R}-\m{u_L} = \m{d_L} - \m{u_r} = \m{e_L}-\m{e_R} \,. \label{eq:chemequ2}
\end{align}
The chemical potentials can be related to an asymmetry in a charge $q$ as 
\begin{align}
  Y_{\Delta q} = q = \frac{T^2}{6s} \left[\sum_{i\in f} q_ig_i \mu_i + 2\sum_{j\in b} q_j g \mu_j\right] = \sum_{i \in SM} \frac{n_i - n_{\overline i}}{s}\,,
\end{align}
where $g_{i,j}$ are the numbers of degrees of freedom of the particle. 
Above the EWPT, charge and hypercharge neutrality result in 
\begin{align}
   Y_{\Delta \mathcal B} = \frac{28}{79} Y_{\Delta(\mathcal B- \mathcal L)}\,,
\end{align}
while charge neutrality and the Bose-Einstein condensate of the Higgs after the EWPT result in 
\begin{align}
   Y_{\Delta \mathcal B} = \frac{12}{37}Y_{\Delta(\mathcal B- \mathcal L)} \,.
\end{align}
\subsection{Domain Walls in the Majoron Model}
In this section, we will shortly discuss the appearance of domain walls in the Majoron model \cite{dw1}. As has been discussed in the previous sections, sphaleron transitions are frequent at high temperatures, $ T \approx \SI{e12}{\giga\electronvolt}$, and explicitly break the $U(1)_{\mathcal L}$ to a residual $Z_3$. 
On the other hand, $\sigma$ obtains its VEV at a slightly lower temperature when sphaleron processes are still active and thereby spontaneously breaks the discrete $Z_3$, resulting in the appearance of stable domain walls. \footnote{{Note that the domain walls in the Majoron model could be rendered unstable by assuming an appropriate effective potential, as discussed in \cite{dw2}.}}
Their energy density can be approximated as $\rho_{DW}\sim \frac{f^2 \sigma}{2t}  \propto T^2$ which is bound to exceed the radiation energy density, $\rho_{rad} = \frac{\pi^2}{30}g^* T^4$ at some time\footnote{Here, $g^*$ is the number of effective degrees of freedom and $t$ is the cosmic time, $t = \frac{1}{T^2}M_{pl}\left( \frac{45}{16\pi^3} \right)^2\frac{1}{\sqrt{g^*} }$ and $M_{pl}$ is the Planck mass.}. For example, domain walls will dominate the energy density around $T \sim \SI{15}{\giga\electronvolt}$ if we assume a reasonable seesaw scale of  $f \approx \SI{e9}{\giga\electronvolt}$. This is in stark contrast to observations and thus requires modifications of the Majoron model in order to stay consistent with cosmology.\footnote{Possible solutions we will not consider here are inverse Majoron models in which the spontaneous symmetry breaking happens when sphaleron transitions seize to be frequent and thus the previous $U(1)_{\mathcal L}$ is restored. \cite{dw1}} 
%Instead, we will focus on extensions of the Majoron model in which the residual discrete symmetries are altered. There are two possible scenarios: Either instanton/sphaleron transitions explicitly break $U(1)_{\mathcal L}$ completely, i.e. there is no discrete symmetry left that can be spontaneously broken, or both residual symmetries are identical so that no spontaneous symmetry breaking of a discrete symmetry takes place.}

\section{Majoron Models without Domain Walls}
\label{sec:nodw}
In this section, we discuss solutions to the domain wall problem and their impact on leptogenesis. As has been pointed out in the previous section, the residual symmetry related to instanton and sphaleron processes is correlated with the anomaly coefficient $\mathcal A_{\lp}$. Since the anomly coefficient depends on the particle content of the model we can easily change $\mathcal A_{\lp}$ by extending the Majoron model by additional right-handed particles \cite{dw1}. 
{If we change the anomly coefficient in a way that prevents the mismatch of the residual discrete symmetries related to instanton processes and SSB at the seesaw scale, domain walls never arise.}
We can envisage two simple domain wall free extensions: If $|\mathcal A_{\lp}| = 1$, sphaleron transitions explicitly break $U(1)_{\mathcal L}$ completely, i.e. there is no discrete symmetry left that can be spontaneously broken. If $|\mathcal A_{\lp}| = 2$, both residual symmetries are equivalent, i.e. no SSB of a discrete symmetry takes place. Thus, models with an altered anomaly coefficient $|\mathcal A_{\lp}| = 1,2$ result in a Majoron model that is free from the appearance of {stable or unstable} domain walls. 
Moreover, the change in the $\left[SU(2)_L\right]^2\times U(1)_{\mathcal L}$ anomaly leaves $A_{\bn}$ unaffected, i.e. $\mathcal A_{\mathcal{B}} \neq \mathcal A_{\mathcal{L}}$, and consequently, instanton and sphaleron processes no longer conserve $\mathcal{B}-\mathcal{L}$ but different combinations of $\mathcal{B}$ and $\mathcal{L}$, depending on $\mathcal A_{\lp}$. 
As a result, the condition between the chemical potentials of SM particles induced by sphaleron transitions being in thermal equilibrium changes, strongly affecting leptogenesis. \\
For simplicity, we will restrict our disussion to extensions using right-handed doublets and triplets \footnote{We want to stress that the models considered here do not exhaust the vast amount of possibilites of domain wall free extensions to the Majoron model.}, 
\begin{align}
  \chi_{R_i} &= \begin{pmatrix} \sfrac{\chi^0_{R_i}}{\sqrt 2} & \chi_{R_i}^+ \\ \chi_{R_i}
  ^- & -\sfrac{\chi^0_{R_i}}{\sqrt 2}   \end{pmatrix} \sim (1,3,0)_1\,,\\
  \eta_{R_i} &= \begin{pmatrix} \eta_{R_i}^0 \\ \eta_{R_i}^- \end{pmatrix} \sim (1,2,-\frac{1}{2})_1\,, 
\end{align}
where $i = 1, ..., N_{\chi_R, \eta_R}$ are generation indices for the new doublets and triplets, respectively. 
In the models with $N_{\eta_R}$ doublets, we introduce the same number of left-handed doublets $\xi_L$ with vanishing lepton number, 
\begin{align}
  \xi_{L_i} &= \begin{pmatrix} \xi_{L_i}^0 \\ \xi_{L_i}^- \end{pmatrix} \sim (1,2,-\frac{1}{2})_0\,, 
\end{align}
in order to cancel gauge anomalies. Moreover, we assume that bare mass terms with the SM leptons are absent and that the doublets acquire a mass at a higher energy scale in order to avoid the appearance of Landau poles due to the running of the $SU(2)_L$ and $U(1)_Y$ gauge couplings.\\
The kinetic and yukawa terms of the Lagrangian are now given by 
\begin{align}
  \lag_{kin}^{new} &= \mathrm{i}\overline{N_R}\slashed\partial N_R + \mathrm{i}\overline{\eta_R}\slashed D\eta_R + \mathrm{i}\overline{\xi_L}\slashed D\xi_L + \mathrm{i}\tr[\overline{\chi_R}\slashed D\chi_R]\,, \\
  \begin{split}
  \lag_{yuk}^{new} &= -y_{H_{ij}}^\nu \overline{L_{L_i}}\tilde HN_{R_j}
 -y_{H_{ij}}^\chi\overline{L_{L_i}}\chi_{R_j}\tilde H - \frac{1}{2} g_{N_{ij}} \overline{N_{R_i}^c}N_{R_j}\sigma \\
   &- \frac{1}{2} g_{\chi_{ij}} \tr[{\overline{\chi_{R_i}^c}\chi_{R_j}}\sigma] + \hc\,, \label{eq:lag_kin}
  \end{split}
\end{align}
where the covariant derivative is given by 
\begin{align}
D_\mu = \partial_\mu -\mathrm{i}\frac{e}{\sin\theta_W\cos\theta_W} Z_\mu (T^3-\sin^2\theta_WQ) - \mathrm{i}eA_\mu Q - \mathrm{i}\frac{g}{\sqrt2}(W_\mu^+T^+ + W_\mu^-T^-) \,.
\end{align}
Here, $T^3$ is the weak isospin of the respective particle, $Q$ is the electric charge and $\theta_W$ is the Weinberg angle. 
{Considering the model with additional triplets, the second term in \eqref{eq:lag_kin} introduces mixing between the triplets and the SM leptons and thereby introduces an additional channel for $\lp$ violation. After SSB at the seesaw scale $f$ and the electroweak scale $v$, the light neutrinos obtain masses via a type I+III seesaw. Restricting ourselves to one generations of triplets, the Dirac and Majorana matrices $m_D$ and $M_R$ are given by 
\begin{align}
  m_D = \begin{pmatrix} vy_{H_1}^\nu & vy_{H_2}^\nu  & vy_{H_3}^\nu \\ vy_{H_1}^\chi & vy_{H_2}^\chi  & vy_{H_3}^\chi  \end{pmatrix}\,,\qquad M_R = \begin{pmatrix} M_R^N & 0 \\ 0 & M_R^\chi\end{pmatrix}\,,
\end{align}
resulting in one massless light neutrino while the other light neutrinos obtain masses as
\begin{align}
  m_\nu^{ij} = - m_D^TM_R^{-1}m_D \approx - \frac{v^2}{2}\left( \frac{y_{H_i}^\nu y_{H_j}^\nu}{M_R^N}+ \frac{y_{H_i}^\chi y_{H_j}^\chi}{M_R^\chi} \right)\,,
\end{align}
where the mass eigenstates of the heavy neutrinos and the triplet are given by $M_R^N = \frac{g_N f}{\sqrt 2}$ and $M_R^\chi = \frac{g_\chi f}{\sqrt 2}$, respectively. Thus, we can easily generate masses for the heavy neutrinos of order of the seesaw scale $f$.
 Lower limits of $M_R^\chi \sim \mathcal{O}(\SI{500}{\giga\electronvolt})$ have been set by Atlas \cite{triplet_atlas} and CMS \cite{triplet_cms}.
}
Moreover, we considered models in which an additional $Z_2$ symmetry is envoked under which the triplets are odd and all other particles even. In this case, the second term in \eqref{eq:lag_kin} is forbidden and no mixing between SM leptons and triplets occurs. 
{In this case, the mass of the triplet is given by $M_R^\chi = \frac{g_\chi f}{\sqrt 2}$ and the light neutrinos obtain a mass via the type I seesaw presented in \ref{sec:majoron_setup}.}
If the {second term in \eqref{eq:lag_kin}} is present, it introduces an additional channel for $\lp$ violation. \\
A summary of the models discussed in the following can be found in Tab. \ref{tab:models}. As we considered models with no more than one triplet, the respective generation index will be dropped from now on. 
The conserved combinations of $\mathcal{B}$ and $\mathcal{L}$ for each model are displayed in the fourth column in Tab. \eqref{tab:models} in terms of $Y_{\mathcal B}$ and $Y_{\mathcal L}$ where $Y_{\Delta \mathcal{B},\mathcal{L}}$ is the respective conserved combination.  
In the fifth column of Tab. \ref{tab:models}, the altered relation between the chemical potentials due to the sphaleron processes being in thermal equilibrium is given. For simplicity, we made the assumption that the right-handed neutrinos are not in thermal equilibrium as is common in leptogenesis scenarios. As the Majoron plays the role of DM, we assume that all processes involving the Majoron are out of thermal equilibrium as well. 
\begin{table}[H]
  \centering
  \begin{tabular}{c c c c c c}
  \toprule
  {Model} & $N_{\chi_R}$& $N_{\eta_R}$ &{$A_{\mathcal L}$}& {$Y_{\Delta(\mathcal{B},\mathcal{L})}$} & {sphaleron}\\
  \midrule 
  {SM}              & $0$& $0$ & $3$ & $Y_{\mathcal B}-Y_{\mathcal L}$  & $2\m{d_L} + \m{u_L}+\m{\nu_L}= 0$\\
  {$1T$}            & $1$& $0$ & $-1$  & $Y_{\mathcal B}+3Y_{\mathcal L}$    & $ \m{d_R}+8\m{d_R} +11\m{e_R}-6\m{e_L}-6\m{\nu_L} = 0$\\
  {$1T + Z_2$}      & $1$& $0$ & $-1$  & $ Y_{\mathcal B}+3Y_{\mathcal L}$    & $\m{d_R}+8\m{d_R} +11\m{e_R}-6\m{e_L}-6\m{\nu_L}= 0$\\
  {$1D$}            & $0$& $1$ & $2$ & $2Y_{\mathcal B}-3Y_{\mathcal L}$   & $\m{d_R} + 8\m{u_R}+8\m{e_R}  - 3\m{e_L}- 3\m{\nu_L} = 0 $\\
  {$2D$}            & $0$& $2$ & $1$ & $Y_{\mathcal B}-3Y_{\mathcal L}$    & $\m{d_R} + 8\m{u_R}+9\m{e_R}  - 4\m{e_L}- 4\m{\nu_L} = 0$\\
  {$4D$}            & $0$& $4$ & $-1$  & $Y_{\mathcal B}+3Y_{\mathcal L}$    & $\m{d_R}+8\m{d_R} +11\m{e_R}-6\m{e_L}-6\m{\nu_L}= 0$\\
  {$5D$}            & $0$& $5$ & $-2$  & $2Y_{\mathcal B}+3Y_{\mathcal L}$   & $\m{d_R}+8\m{d_R} +12\m{e_R}-7\m{e_L}-7\m{\nu_L} = 0 $\\
  {$1D + 1T$}       & $1$& $1$ & $-2$  & $2Y_{\mathcal B}+3Y_{\mathcal L}$   &  $\m{d_R}+8\m{d_R} +12\m{e_R}-7\m{e_L}-7\m{\nu_L} = 0 $\\
  {$1D + 1T + Z_2$} & $1$& $1$ & $-2$  & $2Y_{\mathcal B}+3Y_{\mathcal L}$   &  $\m{d_R}+8\m{d_R} +12\m{e_R}-7\m{e_L}-7\m{\nu_L}= 0 $\\
  \bottomrule
\end{tabular}
\caption{Overview of the models discussed in this paper. $N_{\chi_R, \eta_R}$ are generation indices for the triplet $\chi_R$ and the doublet $\eta_R$, respectively. $A_{\lp}$ is the $[SU(2)_L]^2\times \uell$ anomaly coefficient of the respective model and $Y_{\Delta(\mathcal{B},\mathcal{L})}$ is the corresponding conserved combination of $\bn$ and $\lp$. In the fifth column, the relation between the chemical potentials of SM particles induced by sphaleron processes being in thermal equilibrium is given. }
\label{tab:models}
\end{table}
As we did for the SM, we can relate the baryon asymmetry of the Universe $Y_{\Delta \mathcal{B}}$ to a previous $Y_{\Delta \mathcal{B},\mathcal{L}}$ asymmetry. Besides the changes due to the altered conserved combination of $\bn$ and $\lp$, the conversion rate is changed depending on whether the gauge and yukawa interactions involving $\chi_R$ and $\eta_{R_i}$ are in thermal equlibrium. We assume that the interactions of all generations of doublets $\eta_{R_i}$ are either in thermal equilibrium or out of thermal equilibrium. 
If processes involving $\chi_R$ and $\eta_R$ are in thermal equlibrium, the following relations between the chemical potentials are induced: 
\begin{align}
  \m{W^-} &= \m{\eta_R^-}- \m{\eta_R^0}\,, \\
  \m{W^-} &= \m{\xi_L^-}- \m{\xi_L^0}\,, \\
  \m{W^-} &= \m{\overline{\chi_R^0}} + \m{\chi_R^-}= \m{\chi_R^0} + \m{\overline{\chi_R^+}}\,,\\
  \m{\phi^0} &= \m{\chi^0_R}- \m{\nu_L} = \m{\chi_R^-} - \m{e_L}\,, \label{eq:cond1}\\
  \m{\phi^-} &= \m{\nu_L}- \m{\chi_R^+} = \m{e_L} - \m{\chi_R^0} \,.\label{eq:cond2}
\end{align}
Note that the conditions \eqref{eq:cond1} and \eqref{eq:cond2} only appear in the models with an additional triplet but without $Z_2$ symmetry. Moreover, the chemical potentials for the different generations of doublets are identical, thus we dropped the generation index. 
For simplicity, we assumed that both the gauge and yukawa interactions of the new particles are either in or out of thermal equlibrium while the sphaleron processes are active. In a realistic model, it needs to be thoroughly studied when the respective processes are in thermal equlibrium. Especially the gauge interactions are expected to effectively thermalize the new particles at high temperatures while potentionally falling out of thermal equilibrium at lower temperatures. The details depend strongly on the masses of the new particles and are beyond the scope of this paper.
%\footnote{Note that the doublets are massless in the minimal model presented here and consequently, their kinematics are fully determined by the gauge interactions. Nevertheless, it is possible to extend this model in a way that allows the doublets to become massive, for example by introducing a scalar particle $S$ that obtains a VEV. }\\
Using the conditions given above and those given in \eqref{eq:chemequ1} and \eqref{eq:chemequ2}, relations between $Y_{\Delta \mathcal B}$, $Y_{\Delta \mathcal B, \mathcal L}$, $Y_{\Delta\eta_R^-}$ and $Y_{\Delta\chi_R^0}$ can be found for each model. The results are given in Tab. \ref{tab:no_te} and Tab. \ref{tab:te}. We observe that the change in the anomaly coefficient can have significant effects on the amount of $\lp$ asymmetry that is transferred to a $\bn$ asymmetry.
Given that the additional particles are not in thermal equlibrium, the amount of asymmetry that is transferred is depleted compared to the SM scenario. 
If the additional triplet or doublets are in thermal equilibrium, $Y_{\Delta\eta^-_R}$ and $Y_{\Delta\chi^0_R}$ appear as free parameters in the conversion rate. Consequently, the final $\bn$ asymmetry depends significantly on the details of the specific model. In both cases, a thorough study is necessary to determine whether the new processes result in a washout of the $\lp$ asymmetry or enhance it. Especially the {$1T$} model is expected to be interesting in this regard due to the additional $\lp$ violating interactions compared to the other models. 
\begin{table}[H]
  \centering
  \begin{tabular}{c c c}
  \toprule
  {Model} &  $Y_{\Delta \mathcal B}|_{T>T_{EWPT}}$ & $Y_{\Delta \mathcal B}|_{T<T_{EWPT}} $\\
  \midrule 
  {SM}              & $\frac{28}{79}Y_{\Delta(\mathcal B,\mathcal L)}$ & $\frac{12}{37}Y_{\Delta(\mathcal B,\mathcal L)}$\\
  {$1T$}            & $\frac{76}{679}Y_{\Delta(\mathcal B,\mathcal L)}$  & $\frac{20}{269} Y_{\Delta(\mathcal B,\mathcal L)}$  \\
  {$1T + Z_2$}      & $\frac{76}{679}Y_{\Delta(\mathcal B,\mathcal L)}$  & $\frac{20}{269} Y_{\Delta(\mathcal B,\mathcal L)}$ \\
  {$1D$}            & $\frac{4}{53}Y_{\Delta(\mathcal B,\mathcal L)}$  & $\frac{2}{25}Y_{\Delta(\mathcal B,\mathcal L)}$  \\
  {$2D$}            &$\frac{4}{535}Y_{\Delta(\mathcal B,\mathcal L)}$  & $\frac{8}{246}Y_{\Delta(\mathcal B,\mathcal L)}$   \\
  {$4D$}            &$\frac{76}{679}Y_{\Delta(\mathcal B,\mathcal L)}$  & $\frac{20}{269}Y_{\Delta(\mathcal B,\mathcal L)}$  \\
  {$5D$}            & $\frac{116}{871}Y_{\Delta(\mathcal B,\mathcal L)}$  & $\frac{2}{192}Y_{\Delta(\mathcal B,\mathcal L)}$ \\
  {$1D + 1T$}       & $\frac{116}{871}Y_{\Delta(\mathcal B,\mathcal L)}$  & $\frac{2}{19}Y_{\Delta(\mathcal B,\mathcal L)}$  \\
  {$1D + 1T + Z_2$} & $\frac{116}{871}Y_{\Delta(\mathcal B,\mathcal L)}$  & $\frac{2}{19}Y_{\Delta(\mathcal B,\mathcal L)}$  \\
  \bottomrule
\end{tabular}
\caption{Relations between $Y_{\Delta \mathcal B}$ and $Y_{\Delta(\mathcal B,\mathcal L)}$ if $\chi_R$ and $\eta_{R_i}$ are not in thermal equilibrium. In the second column, the conversion rate for temperatures above the EWPT is given, while the third row displays the conversion rate for temperatures below the EWPT. }
\label{tab:no_te}
\end{table}

\begin{table}[H]
  \centering
  \begin{tabular}{c c c}
  \toprule
  {Model} &  $Y_{\Delta \mathcal B}|_{T>T_{EWPT}}$ & $Y_{\Delta \mathcal B}|_{T<T_{EWPT}}$\\
  \midrule 
  {SM}              & $\frac{28}{79}Y_{\Delta(\mathcal B,\mathcal L)}$ & $\frac{12}{37}Y_{\Delta(\mathcal B,\mathcal L)}$\\
  {$1T$}            & $\frac{4}{67} Y_{\Delta(\mathcal B,\mathcal L)}$  & $\frac{5}{92}Y_{\Delta(\mathcal B,\mathcal L)}$  \\
  {$1T + Z_2$}      & $ \frac{4}{679}\left( 19\yd{\bn,\lp} - 288\yd{\chi^0_R} \right) $  & $ \frac{20}{269}\left( \yd{\bn,\lp} - 9\yd{\chi^0_R} \right) $ \\
  {$1D$}            & $ \frac{2}{258}\left( 22\yd{\bn,\lp} + 405\yd{\eta^0_R} + 9\yd{\xi^0_L} \right) $  & $ -\frac{2}{299}\left(  4\yd{\bn,\lp}  - 1269 \yd{\xi^-_R} + 1197\yd{\xi^0_L}\right) $  \\
  {$2D$}            & $ \frac{4}{535}\left( \yd{\bn,\lp} + 81\yd{\eta^0_R} + 45 \yd{\xi^0_L} \right) $  & $ -\frac{2}{275}\left( 2\yd{\bn,\lp} -549\yd{\xi^-_R} + 477\yd{\xi^0_L} \right) $\\
  {$4D$}            & $ \frac{2}{679}\left(38\yd{\bn\lp} -2385\yd{\eta^0_R} +351\yd{\xi^0_L} \right) $  & $ -\frac{2}{203}\left(  2\yd{\bn,\lp} +1107\yd{\xi^-_R} - 963\yd{\xi^0_L} \right) $  \\
  {$5D$}            &  $ \frac{2}{871}\left( 58\yd{\bn,\lp} -4455\yd{\eta^0_R} + 765\yd{\xi^0_L} \right) $  & $-\frac{2}{155}\left(  4\yd{\bn,\lp}  +2043\yd{\xi^-_R} - 1683\yd{\xi^0_L}\right) $\\
  {$1D + 1T$}       &  $ \frac{1}{92}\left( \yd{\bn,\lp} -1368\yd{\xi^-_R} +1386\yd{\xi^0_L} \right) $  & $ \frac{-2}{37}\left( 2\yd{\bn,\lp} +837\yd{\xi^-_R} - 801\yd{\xi^0_L} \right) $ \\

  {$1D + 1T + Z_2$} & $\frac{2}{871}\left(58Y_{\Delta(\mathcal B,\mathcal L)}+891Y_{\Delta\eta^0_R}\right.$  & $-\frac{2}{155}\left(4Y_{\Delta(\mathcal B,\mathcal L)}-108Y_{\Delta\chi^0_R}\right.$  \\

  {} & $\left.+10647Y_{\Delta\xi^-_R}-10494Y_{\Delta\xi^0_R}\right)$  & $\left.+1755Y_{\Delta\xi^-_R}-1683Y_{\Delta\xi^0_R}\right)$  \\
  \bottomrule
\end{tabular}
\caption{Relations between $Y_{\Delta \mathcal B}$ and $Y_{\Delta(\mathcal B,\mathcal L)}$ if $\chi_R$ and $\eta_{R_i}$ are in thermal equilibrium. In the second column, the conversion rate for temperatures above the EWPT is given, while the third row displays the conversion rate for temperatures below the EWPT.}
\label{tab:te}
\end{table}

\section{Summary}
In this work, we showed that the domain wall problem in the Majoron model can be avoided by introducing new particles with non-trivial $[SU(2)_L]^2\times\uell$ quantum numbers.\\
In general, instanton processes break the $\uell$ symmetry of the SM to a residual $Z_3$, while in the Majoron model, the VEV of a complex scalar $\sigma$ breaks the $\uell$ symmetry to a residual $Z_2$. The mismatch of discrete symmetries results in the appearance of highly undesired cosmological domain walls. We considered extensions of the Majoron model that avoid the domain wall problem, in particular extensions by right-handed doublets $\eta_R$ and triplets $\chi_R$ that result in a domain-wall-free model. These new particles have an interesting impact on leptogenesis as they change the anomaly factor $\mathcal{A}_{\lp}$ and thereby the residual discrete symmetry related to the instantion processes. \\
Since the extensions leave the anomaly factor $\mathcal{A}_{\bn}$ unchanged while changing $\mathcal{A}_{\lp}$, the sphaleron processes that convert an initial $\lp$ asymmetry to a $\bn$ asymmetry conserve different combinations of $\lp$ and $\bn$ compared to the standard scenario. Consequently, the $\lp$ to $\bn$ conversion rate is changed. If the doublets and triplets are not in thermal equilibrium, 
we find that the conversion rate can be in a range from slightly smaller to significantly larger compared to the conventional leptogenesis mechanism, depending on which specific model is considered. If the new particles are in thermal equilibrium, the asymmetries in the particle number densities $Y_{\eta_R^-}$ and $Y_{\chi_R^0}$ appear as free parameters. Consequently, the conversion rate depends on the details of the model that govern the evolution of $\chi_R$ and $\eta_R$. Besides changing the conversion rate, the additional particles can also have an impact on the inital $\lp$ asymmetry due to additional processes that may enhance the asymmetry or increase the washout. These details depend on the specifics of the model and will be the subject of future works. 

\section*{Acknowledgments}
I would like to thank Heinrich Päs, Dominik Hellmann, Rigo Bause, Mathias Becker, Tim Höhne and Tom Steudtner for useful discussions. This work was supported by the \textit{Studienstiftung des deutschen Volkes}.

%%%%%%%%%%%%%%%%%%%%%%%%%%%%%%%%%%%%%

%\bibliographystyle{unsrtdin}
%\bibliography{references}

%\printbibliography

\end{document}